\documentclass[prc,aps,amsfonts,amsmath,superscriptaddress,showpacs,floatfix,twocolumn]{revtex4}
\usepackage{graphicx} 
\usepackage{epsfig} 
\usepackage{rotating}
\usepackage{mwe}    
\usepackage{subfig}

\begin{document}

\title{Spin-orbit coupling rule in bound fermions systems}

\author{J.-P. Ebran}
\affiliation{CEA,DAM,DIF, F-91297 Arpajon, France}
\author{E. Khan}
\affiliation{Institut de Physique Nucl\'eaire, Universit\'e Paris-Sud, IN2P3-CNRS, 
F-91406 Orsay Cedex, France}
\author{A. Mutschler}
\affiliation{Institut de Physique Nucl\'eaire, Universit\'e Paris-Sud, IN2P3-CNRS, 
F-91406 Orsay Cedex, France}
\affiliation{Grand Acc\'el\'erateur National d'Ions Lourds (GANIL),
CEA/DSM-CNRS/IN2P3, Bvd Henri Becquerel, 14076 Caen, France}
\author{D. Vretenar}
\affiliation{Physics Department, Faculty of Science, University of
Zagreb, 10000 Zagreb, Croatia}

\begin{abstract} 
Spin-orbit coupling characterizes quantum systems such as atoms, nuclei, 
hypernuclei, quarkonia, etc., and is essential for understanding their 
spectroscopic properties. Depending on the system, the effect of spin-orbit 
coupling on shell structure is large in nuclei, small in quarkonia, perturbative 
in atoms. In the standard non-relativistic reduction of the single-particle
Dirac equation, we derive a universal rule for the relative magnitude of the spin-orbit 
effect that applies to very different quantum systems, regardless of whether 
the spin-orbit coupling originates from the strong or electromagnetic interaction. 
It is shown that in nuclei the near equality of the mass of the nucleon and the 
difference between the large repulsive and attractive potentials explains the 
fact that spin-orbit splittings are comparable to the energy spacing between 
major shells. For a specific ratio between the particle mass and the effective 
potential whose gradient determines the spin-orbit force, we predict the occurrence 
of giant spin-orbit energy splittings that dominate the single-particle excitation spectrum.
\end{abstract}


\date{\today}

\maketitle

\section{Introduction}

Spin-orbit coupling, known since the early days of quantum mechanics,
is among the most studied effects related to the spin of a particle. After
nearly a century, spin-orbit effects continue to provide a basis for a
variety of new phenomena in diverse fields of quantum physics, such as 
spintronics \cite{spintronics}, topological insulators
\cite{topological}, cold atomic gases \cite{quantumgases}, atomic
nuclei far from stability \cite{SP.08}, etc. In nuclear physics, in
particular, the strong coupling between the orbital angular momentum
and spin of a nucleon accounts for the empirical magic numbers and
shell gaps. The energy spacings between spin-orbit partner states 
can be as large as the gaps between major shells in
atomic nuclei, but the spin-orbit splittings can also be considerably
reduced in short-lived nuclei with extreme isospin values
\cite{SP.08,LVPR.98,Burg.14}. Starting from the nuclear physics case,
we investigate the ratio between the major energy spacings of levels
characterized by principal single-particle quantum numbers and the
energy splitting of spin-orbit partner states in systems of bound
fermions. This ratio can differ, of course, by orders of magnitude in
systems in which the spin-orbit coupling originates from different
underlying interactions, e.g. the strong effective nuclear force or
the electromagnetic interaction. It is, therefore, interesting to try
to find a common characteristic for the spin-orbit splitting in
systems as diverse as atoms, nuclei, hypernuclei, quarkonia, etc. 

\section{Spin-orbit coupling in atomic nuclei}

In the nuclear relativistic mean-field framework \cite{Duerr,SW.86,SW.97} 
a nucleus is considered as a system of independent nucleons moving in 
local self-consistent scalar and vector potentials. The single-nucleon dynamics is 
governed by the Dirac equation: 
\begin{equation}
\label{Dirac}
\left [ {\vec \alpha} \cdot {\vec p} + V + \beta (m+S) \right ] \psi_i = E_i \psi_i
\end{equation}
where $\psi_i$ denotes the Dirac spinor: 
\begin{equation}
\left(
\begin{array}[c]{c}
\phi_i\\
\chi_i
\end{array}
\right)
\end{equation} 
for the {\em i}-th nucleon. For simplicity we only consider spherical nuclei and assume 
time-reversal symmetry (pairwise occupied states with Kramers degeneracy), which 
ensures that the only non-vanishing components of the vector fields are the time-like 
ones and thus there is no net contribution from nucleon currents. The local 
vector $V$ and scalar $S$ potentials are uniquely determined by the actual 
nucleon density and scalar 
density of a given nucleus, respectively. In the ground state A nucleons occupy the lowest 
single-nucleon orbitals, determined self-consistently by the iterative solution 
of the Dirac equation (\ref{Dirac}). If one expresses 
the single-nucleon energy as $E_i = m + \varepsilon_i$, 
where $m$ is the nucleon mass, and 
rewrites the Dirac equation as a system of two equations for $\phi_i$ and $\chi_i$, then, 
noticing that for bound states $\varepsilon_i << m$, 
\begin{equation}
\chi_i \approx {1\over 2 {\cal M}(r)} ({\vec \sigma} \cdot {\vec p}) ~\phi_i 
\end{equation} 
to order $\varepsilon_i / m$, and 
\begin{equation}
{\cal M}({r})\equiv m+\frac{1}{2}\left(S({r})-V({r})\right) \;.
\label{eq:meff}
\end{equation}
The equation for the upper component $\phi_i$ of the Dirac spinor reduces to the 
Schr\" odinger-like form \cite{bib82,rei89} 
\begin{equation}
\label{Sch}
\left [ {\vec p} {1 \over 2 {\cal M}(r)} {\vec p} + U(r) +  V^{LS}(r) \right ] \phi_i = \varepsilon_i   \phi_i
\end{equation} 
for a nucleon with effective mass ${\cal M}({r})$ in the potential
$U(r) \equiv V(r) + S(r)$, and with the 
spin-orbit potential: 
\begin{equation}
\label{vls1}
V^{LS} \equiv {1 \over 2 {\cal M}^2 (r)} {1\over r} {d\over dr} \left ( V(r) - S(r) \right ) {\vec l} \cdot
{\vec s} \;.
\end{equation} 
The spin-orbit coupling plays a crucial role in nuclear structure, and its   
inclusion in the effective single-nucleon potential is essential 
to reproduce the empirical magic numbers. The relativistic mean-field 
framework, in particular, naturally includes the nucleon spin degree of freedom, and 
the resulting spin-orbit potential emerges automatically with the empirical 
strength \cite{FS.00}. The nuclear spin-orbit potential originates from the 
difference between two large fields: the vector potential $V$ (short-range repulsion) 
with typical strength of $\approx 350$ MeV, and the 
scalar potential $S$ (medium-range attraction), typically of the order of 
$- 400$ MeV in nucleonic matter and finite nuclei. 
In the context of in-medium QCD sum rules \cite{inmediumqcdsumrules}, 
the strong scalar and vector mean fields experienced by nucleons can 
be associated with the leading density dependence 
of the chiral (quark) condensate, $\langle\bar{q}q\rangle$, 
and the quark density $\langle q^\dagger q\rangle$. In the mean-field phenomenology 
the sum of these two fields $V+S \approx - 50$ MeV provides 
the confining potential that binds the nucleons in a nucleus, whereas the 
large difference $V-S \approx 750$ MeV determines the pronounced energy 
spacings between spin-orbit partner states in finite nuclei, of the 
order of several MeV \cite{Duerr,SW.86,SW.97,FS.00,LVR.98,vre05}. 
A puzzling coincidence that we would like to explore is that the 
largest spin-orbit splittings for intruder states  are comparable in magnitude 
to the energy gaps between major shells of the nuclear potential. 
 
The aim of this study is to evaluate the typical ratio between the 
energy spacings of levels characterized by principal single-particle 
quantum numbers and the energy splitting of spin-orbit partner 
states (fine structure). Even though the values 
of this ratio span orders of magnitude for different bound quantum 
systems, we will show that it is basically governed by two 
quantities that characterize a given system, irrespective whether the 
binding and spin-orbit potentials originate from the strong (nuclei) or 
electromagnetic (atoms) interactions. 

In the nuclear case the interaction is of short range and the self-consistent 
potentials display a spatial distribution that corresponds to the actual 
single-nucleon density. 
The expression for the spin-orbit potential Eq.~({\ref{vls1}) can, therefore, be 
rewritten in the following form:
\begin{equation}
V^{LS}\simeq F(r)~\frac{\rho^\prime(r)}{2\rho(r) r} ~ \vec{l}\cdot \vec{s}\; ,
\label{lsred}
\end{equation}
where
\begin{equation}
F(r)\equiv{{V(r)-S(r)} \over {\left [ m-\frac{1}{2} (V(r)-S(r))\right ]^2}}\; ,
\label{F}
\end{equation}
and $\rho$(r) denotes the self-consistent ground-state density 
of a nucleus with A nucleons. 
For a typical approximation of the single-nucleon potential, such as the 
harmonic oscillator, or the more realistic Woods-Saxon potential, 
one can show \cite{BM}:
\begin{equation}
<\frac{\rho'(r)}{2\rho(r) r}>\simeq - \frac{1}{R_0^2}  
\label{rad}
\end{equation}
where $R_0 = r_0 A^{1/3}$, $r_0 \approx 1.2$ fm 
and, together with $< \vec{l}\cdot \vec{s} > = l/2$ for $j=l+1/2$, and 
$< \vec{l}\cdot \vec{s} > = - 1/2 (l+1)$ for $j=l-1/2$, the energy spacing 
between spin-orbit partner states can be approximated by:
\begin{equation}
\label{delta_vls}
| \Delta < V^{LS} > |    
\approx  F {l\hbar^2 \over R_0^2} 
\end{equation}
Let us consider the ratio between the major energy spacings and the 
spin-orbit splitting. For the harmonic oscillator potential one finds \cite{rs}
\begin{equation}
\hbar \omega_0 = {\hbar \over R_0} \sqrt{ -2 U_0 \over m}\;, 
\end{equation}
where, in our case, the depth of the potential is $U_0 \equiv U(r=0) = V(0) + S(0)$. Therefore, 
\begin{equation}
\label{x_ratio}
x\equiv\frac{\hbar\omega_0}{|\Delta < V^{LS}>|} = K \bigg|\eta-1+\frac{1}{4\eta}\bigg| \;, 
\end{equation}
where $K =  \sqrt{- 2 m U_0} R_0 /l \hbar$, and 
\begin{equation}
\eta\equiv\frac{m}{V-S} \;.
\label{eta}
\end{equation}
K is typically of the order $1-5$ for l $\ge$ 3 (corresponding to the
magic spin-orbit gaps). Since for the nucleon mass $m \approx 940$ MeV
and $V - S \approx 750$ MeV: $\eta = 1.25$, it follows from Eq.
(\ref{x_ratio}) that for the nuclear system the ratio $x$ is of the
order $1 - 5$, that is, in nuclei the energy splitting between
spin-orbit partner states is comparable in magnitude to the spacings
between major oscillator shells. This is because of the near equality
of the mass $m$ and the potential $V-S$ in nuclei.  It should be noted
than in the case of the pseudo-spin symmetry \cite{gin97} (V=-S), Eq.
(\ref{eta}) yields $\eta$=m/2V. An aspect that can be generalised to
different quantum systems is the specific functional dependence of $x$
on the ratio of the particle mass $m$ and the effective potential
whose gradient determines the spin-orbit force. 

\section{Other systems of bound fermions}

In the case of atomic systems the binding of an electron is determined
by the Coulomb potential $V(r) = - Z\alpha /r$, where $Z$ is the
charge of the nucleus and the fine-structure constant $\alpha =
1/137$. One can again perform the non-relativistic reduction 
of the Dirac equation for the electron (Eqs. (\ref{Dirac}) - (\ref{Sch})), 
and the resulting spin-orbit 
potential reads:
\begin{equation}
\label{vls2}
V^{LS} = {1 \over 2 {\cal M}^2 (r)} {1\over r} {d V(r)\over dr} ~{\vec l} \cdot {\vec \sigma} \;,
\end{equation} 
with $2{\cal M}({r})\equiv 2m-V({r})$. In this case, of course, $U(r)$
in Eq.~(\ref{Sch}) is just the Coulomb potential. The energy spacing
between successive levels with different principal quantum number $n$
is proportional to $\alpha^2$, whereas the first-order spin-orbit
splitting is $\sim \alpha^4$. The ratio between the principal energy
spacings and the spin-orbit splittings (fine structure) is much larger
than in the nuclear case, that is $\sim 1/\alpha^2 \approx 10^4$,
known from the early seminal work of Sommerfeld \cite{som}. The
interesting fact is that, starting from Eq.~(\ref{vls2}), this ratio
can again be expressed with the same functional dependence on $\eta$
as in Eq.~(\ref{x_ratio}). $\eta = m/V$ is now negative and, with
$m=0.5$ MeV and $V(r_0) = -2.72 \times 10^{-5}$ MeV for the hydrogen
atom and the Bohr's radius $r_0$, it follows that in the atomic case
the characteristic value is $\eta \sim -1/\alpha^2 \approx -2\times
10^{4}$. For large absolute values of $\eta$, the expression
Eq.~(\ref{x_ratio}) reduces to x $\sim 1/\alpha^2 \approx 10^4$, in
agreement with the empirical value quoted above. The fine structure of
atomic spectra thus becomes a limit of the spin-orbit rule
(\ref{x_ratio}). It should be noted the validity of the spin-orbit
rule in Coulomb-like systems is due to the 1/r behavior of the
potential. The spin-orbit rules also applies to the case of ions
having Z protons: in these systems the fine structure is known to behave as
Z$^2\alpha^2$. In the present approach the Z$^2$ factor comes from the
Z one of Eq. (\ref{eta}) and the Z one from the r$_0$/Z typical size
of the ion.

Figure \ref{fig:x} displays the quantity representing the spin-orbit
rule:
\begin{equation}
x\simeq \bigg| \eta-1+\frac{1}{4\eta} \bigg|
\label{x2}
\end{equation}
as a function of the ratio $\eta$ between the mass of the particle and
the effective potential that determines the spin-orbit force in a 
given quantum system. As shown above, for nuclei $\eta$ is slightly 
larger than one and, depending on the specific orbital, $x$ lies in
the interval $1-5$. In atoms $\eta$ is negative and of the order of
$10^{4}$ and, therefore, the characteristic value of $x$ is $\sim
1/\alpha^2 \approx 10^4$. 

The main results of the spin-orbit rule quantatively apply to nuclei
and atoms. It may be relevant to test it in a few other quantum
systems although a more qualitative agreement is expected, due to 
additional effects which would not be included in the present
approach. For instance, experimental evidence indicates that in
$\Lambda$ hypernuclei the $\Lambda$-nuclear spin-orbit interaction is
very weak compared to the strong spin-orbit interaction in ordinary
nuclei \cite{hypspinorbit,hypspinorbit3}. Available data show that the
energy spacings between $\Lambda$ spin-orbit partner states are of the
order of $\approx 100$ keV, although some microscopic models also
predict 2 MeV values in some cases \cite{son09}. Even though a
quantitative study of the smallness of the $\Lambda$ spin-orbit
interaction necessitates a rather involved analysis based on in-medium
chiral SU(3) dynamics \cite{Finellihyper}, one can qualitatively
understand the reduction of the spin-orbit splitting in $\Lambda$
hypernuclei already at the relativistic mean-field level
\cite{QCD-Lambda,MJ.94}. A finite-density QCD sum-rule analysis 
\cite{inmediumqcdsumrules,QCD-Lambda} indicates that, when compared 
the strong scalar and vector mean fields experienced by nucleons in
ordinary nuclei, the corresponding self-energies of a $\Lambda$
hyperon are reduced by a factor $\approx 0.4 - 0.5$, and even smaller
if corrections from in-medium condensates of higher dimensions are
taken into account. With $m_\Lambda = 1.12$ GeV, this means that
$\eta$ in Eq.~(\ref{eta}) is of the order $3-4$, leading to larger x
value than in nuclei, in qualitative agreement with empirical values
of x in such systems. Since the energy spacing between major
oscillator shells in $\Lambda$ hypernuclei (for instance,
$^{13}_\Lambda$C, $^{16}_\Lambda$O, $^{40}_\Lambda$Ca), is of the
order of $\approx 10$ MeV \cite{hypspinorbit}, the empirical ratio $x$
ranges from a few units to a few dozens. Our mean-field estimate is
closer to the small values of x but, as shown in Ref.~\cite{MJ.94}, a
considerable additional reduction of the splitting between $\Lambda$
spin-orbit partner states arises as the effect of the hypernuclear 
tensor coupling. As explained above, a quantitative explanation of the
small spin-orbit splitting in hypernuclei must include
beyond-mean-field effects. 

\begin{figure}[]  
      {\includegraphics[width=0.57\textwidth]{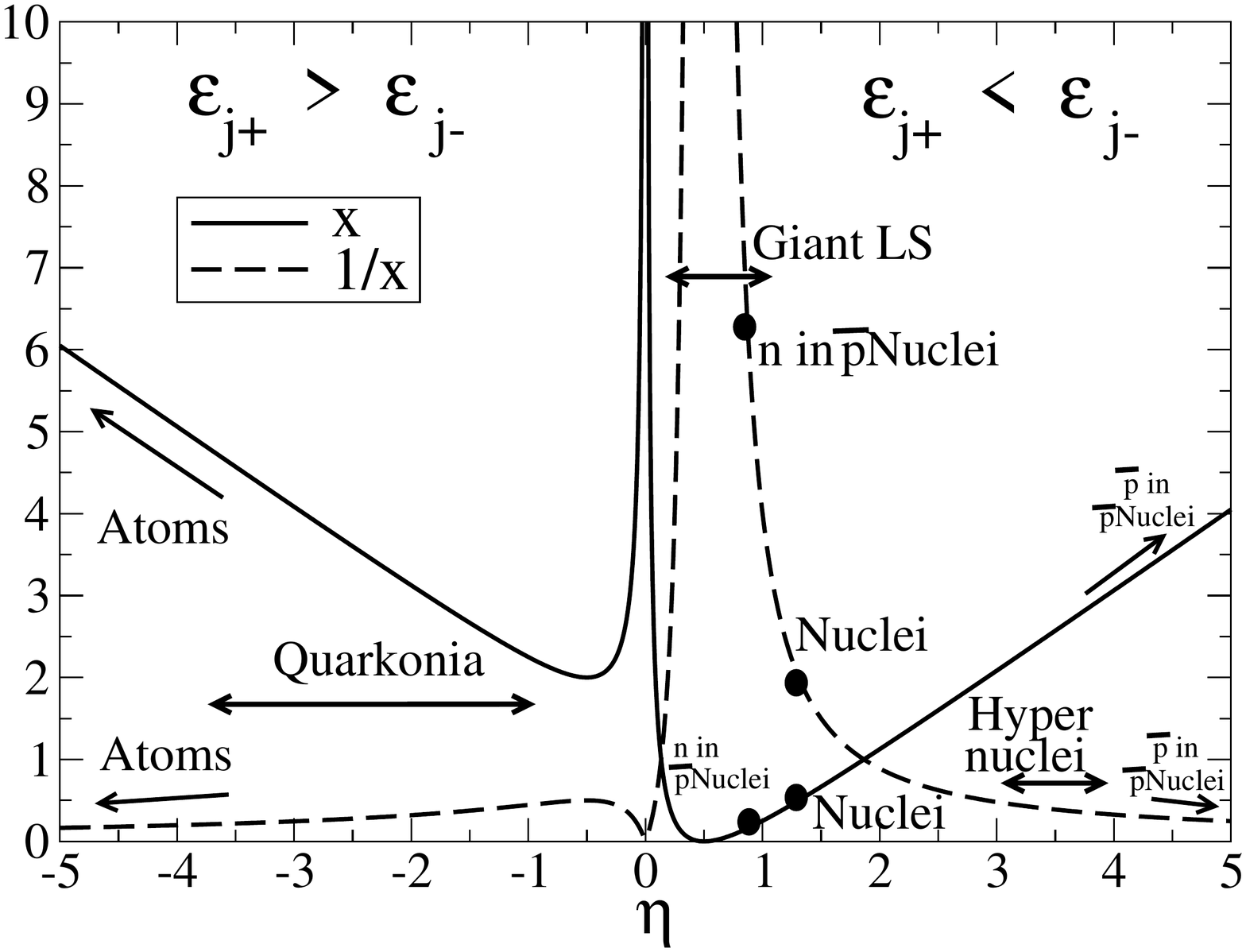}
    }
    \caption{The ratio between the principal energy 
spacings and the spin-orbit splittings (fine structure) Eq.~(\ref{x2}),
as a function of the ratio $\eta$ between the mass of the particle and 
the effective potential that determines the spin-orbit force in a 
given quantum system.}    
    \label{fig:x}
  \end{figure}

An interesting example of quantum systems that are governed by the
strong interaction but exhibit negative values of $\eta$, are
quarkonia such as, for instance, charmonium $\bar{c} c$ and
bottomonium $\bar{b} b$. The center-of-mass is different, compared to
the present one-body approach, but this only generates a global
scaling factor of 2 on the energy positions, due to the reduced mass.
Since we are dealing with order of magnitudes, and most importantly
with energies ratios, it is relevant to check if the present approach
applies in the case of quarkonia. The use of potential models for
these system can be justified by the fact that the bottom and charm
masses are large in comparison to the typical hadronic scale of QCD.
Most phenomenological approaches to the dynamics of two heavy quarks
interacting through a potential are variants of the Cornell model
\cite{cor}, which consists of a superposition of the
one-gluon-exchange that leads to a Coulomb-like attractive vector
potential at short distances:
\begin{equation}
V=-\frac{4}{3}\frac{\alpha_s(r)}{r} \;,
\label{cornell}
\end{equation} 
plus a scalar linear confining potential $S= \sigma r$, with $ \sigma
\approx 0.18$ GeV$^{2}$. Therefore, $V-S$ is negative in the
quarkonia case. The masses of the $c$ quark $b$ quark are $m_c
\approx 1.27$ GeV and $m_b \approx 4.2$ GeV, respectively. At radial
distances that correspond to the mean-square radii of the quarkonia
states the Coulomb-like attractive vector potential
Eq.~(\ref{cornell}), with a depth of the order of $- 1$ GeV, dominates 
over the scalar potential \cite{Buch81}. 
The mass of the lowest charmonium state is $\approx 3$
GeV, while that of the lowest bottomonium state is $\approx 9.5$ GeV.
The energy spacing between $1S$ and $2S$ states is of the order of
$600$ MeV for charmonia, and $560$ MeV for bottomonia
\cite{Brambilla,pdg}. The fine splittings between $1P_J$ states are 
less than 100 MeV for the charmonium ($\Delta M_{21} = 45.6 \pm 0.2$
MeV and $\Delta M_{10} = 95.3 \pm 0.4$ MeV), and less than 40 MeV for
the bottomonium ($\Delta M_{21} = 19.4 \pm 0.4$ MeV and $\Delta M_{10}
= 33.3 \pm 0.5$ MeV). This means that, in the case of quarkonia, the
empirical ratio between the energy spacings characterized by the
principal quantum numbers and the fine spin-orbit splittings is of the
order of 5 to 10, in qualitative agreement with Fig. 1.

The particular dependence of this ratio on $\eta$, shown in
Fig.~\ref{fig:x}, displays interesting features. For positive values
of $\eta$, states for which the orbital angular momentum and spin are
aligned are found at lower energy with respect to states for which the
orbital angular momentum and spin are anti-aligned (nuclei,
hypernuclei), whereas the opposite energy ordering is found for
negative $\eta$ (atoms, quarkonia).  The ratio $x$ (Eq.~(\ref{x2}))
diverges at $\eta=0$, that is, in the limit of massless particle.  

An interesting situation is found in the vicinity of $\eta$=1/2, for
which $x=0$. More precisely, $\eta$ values close to 1/2 are in the
validity domain of the approximation performed to derive the present
spin-orbit rule. This occurs when the mass of the particle is close to
$(V-S)/2$, and the energy spacings between states are characterized by
very large spin-orbit coupling (giant LS). Such states could be
obtained in particular cases for which one would be able to choose the
strength of the effective potential whose gradient determines the
spin-orbit force. In a hypothetical case this limit is approached when
the effective potential becomes very deep as it would occur, for
instance, for bound antibaryon-nuclear systems which, in addition to
ordinary nucleons, contain antibaryons ($\bar{B} = \bar{p},
\bar{\Lambda}, \ldots$) (cf. Figs. 10 and 21 of Ref.~ \cite{gre}). Two
cases can be considered. For the antibaryon spectrum the spin-orbit
splitting remains small, of about few hundreds keV \cite{zho03}. The
corresponding V-S potential is smaller than in the nucleon case
because V has the opposite sign in the case of antibaryons
\cite{zho03} and, therefore, the value of $\eta$ becomes larger (Fig.
\ref{fig:x}). This is both the case for the antiproton and
anti-$\Lambda$ nuclei \cite{son09,son11}: in the last case a good
quantitative agreement is found for the relative amplitude of the
spin-orbit splitting predicted by the spin-orbit rule (x$\sim$ 40).

The case of nucleons in antibaryon-nuclei is closer to
the giant LS state: the gradient of the potential obtained from
self-consistent calculations increases and the spin-orbit splitting
not only gets larger, but displays a pronounced increase with respect
to the spacing between major shells (Fig. \ref{fig:x}). However, such
systems have not been observed and this is just an illustrative
example, with antibaryon-nucleus potentials subject to large
uncertainties \cite{gre}. A similar effect, although much weaker, is
predicted for single-nucleon spectra in $\Lambda$-hypernuclei
\cite{VPLR.98}. Even though the presence of a $\Lambda$ induces only a
fractional change in the central mean-field potential, through a
purely relativistic effect it increases the spin-orbit term in the
surface region, providing additional binding for the outermost
neutrons. To emphasize this region of giant spin-orbit splittings, in
Fig.~\ref{fig:x} we also plot the inverse of $x$. For negative values
of $\eta$ the ratio $x$ does not vanish, but again one finds a region
of small absolute values of $\eta$ for which the splitting between
spin-orbit partner states is comparable in magnitude to the energy
spacing between successive levels with different principal quantum
number. 

\section{Summary}

In conclusion, we have investigated an interesting effect of spin-orbit
coupling in systems of bound fermions (electrons, quarks, nucleons,
hyperons). By performing the usual non-relativistic reduction of the
Dirac equation, the single-particle mean-field equation takes a Schr\"
odinger-like form which, in addition to the confining potential,
exhibits the spin-orbit potential explicitly. Starting from the
relativistic mean-field approximation in the nuclear case, we have
derived an analytic expression for the ratio between the
energy spacings characterized by principal single-particle quantum
numbers and the energy splitting of spin-orbit partner states. This
quantity explicitly depends on the ratio of the particle mass and the 
effective potential whose gradient determines the spin-orbit force
(cf. Eq.~(\ref{x2})). In nuclei the near equality of the mass of the
nucleon and the difference between the large repulsive vector and
attractive scalar potentials explains the fact that the spin-orbit 
splittings are comparable to the energy spacing between major
oscillator shells. The same universal functional form also applies to
other bound quantum systems, regardless of whether the spin-orbit
potential originates from the strong or electromagnetic interaction,
and for which the ratio between the principal energy spacings and the
spin-orbit splittings is orders of magnitude larger than in the
nuclear case. For a given particle mass this ratio could be
altered, in principle, by modifying the potential that generates the
spin-orbit coupling effect. When this ratio is $\eta \approx 1/2$, our
study predicts the occurrence of giant spin-orbit splittings, that is,
a single-particle excitation spectrum dominated by large energy
spacings between spin-orbit partner states.

The extension of the derivation of the spin-orbit rule to exotic
nuclei, where the densities are more diffuse shall be undertaken in a
forthcoming work. This includes halo nuclei involving coupling to 
the continuum \cite{men05,men96,pos97,men98}. It would also be
relevant to derive a pseudo-spin rule, providing the typical magnitude
of the related degeneracy raising due to the breaking of the
pseudo-spin symmetry \cite{gin97,lia15}, with respect to the main shell
structure.

\bigskip 

This work was supported by the Institut Universitaire de
France. The authors thank Nguyen Van Giai, Tamara Nik\v si\'c, Peter
Ring and Olivier Sorlin for reading the manuscript and valuable
discussions.

\end{document}